\documentstyle[12pt]{article}
\newcommand{\simgt}{\,\rlap{\lower 3.5 pt \hbox{$\mathchar \sim$}} \raise 1pt
 \hbox {$>$}\,}
\newcommand{\simlt}{\,\rlap{\lower 3.5 pt \hbox{$\mathchar \sim$}} \raise 1pt
 \hbox {$<$}\,}
\catcode`@=11
\newcount\@tempcntc
\def\@citex[#1]#2{\if@filesw\immediate\write\@auxout{\string\citation{#2}}\fi
  \@tempcnta\z@\@tempcntb\m@ne\def\@citea{}\@cite{\@for\@citeb:=#2\do
    {\@ifundefined
       {b@\@citeb}{\@citeo\@tempcntb\m@ne\@citea\def\@citea{,}{\bf ?}\@warning
       {Citation `\@citeb' on page \thepage \space undefined}}%
    {\setbox\z@\hbox{\global\@tempcntc0\csname b@\@citeb\endcsname\relax}%
     \ifnum\@tempcntc=\z@ \@citeo\@tempcntb\m@ne
       \@citea\def\@citea{,}\hbox{\csname b@\@citeb\endcsname}%
     \else
      \advance\@tempcntb\@ne
      \ifnum\@tempcntb=\@tempcntc
      \else\advance\@tempcntb\m@ne\@citeo
      \@tempcnta\@tempcntc\@tempcntb\@tempcntc\fi\fi}}\@citeo}{#1}}
\def\@citeo{\ifnum\@tempcnta>\@tempcntb\else\@citea\def\@citea{,}%
  \ifnum\@tempcnta=\@tempcntb\the\@tempcnta\else
   {\advance\@tempcnta\@ne\ifnum\@tempcnta=\@tempcntb \else \def\@citea{--}\fi
    \advance\@tempcnta\m@ne\the\@tempcnta\@citea\the\@tempcntb}\fi\fi}
\catcode`@=12
\def\theequation{\arabic{section}.\arabic{equation}} 

\begin{document}

\begin{flushright}
RAL-TR/96--002\\
MPI/PhT/96--005\\
February 1996
\end{flushright}
\vskip1.cm
\begin{center}
{\LARGE{\bf Mixing Renormalization in Majorana}}\\[0.3cm] 
{\LARGE{\bf Neutrino Theories}}\\[2.5cm]
{\large Bernd A.\ Kniehl}$^a$\footnotemark[1]{\large  ~and Apostolos 
Pilaftsis}$^b$\footnote[1]{E-mail addresses: kniehl@vms.mppmu.mpg.de,
pilaftsis@v2.rl.ac.uk}\\[0.4cm]
{\em $^a$Max-Planck-Institut f\"ur Physik, F\"ohringer Ring 6, 
80805 Munich, Germany}\\[0.2cm]
{\em $^b$Rutherford Appleton Laboratory, Chilton, Didcot, Oxon, OX11 0QX, 
UK}
\end{center}
\vskip2.5cm
\centerline{\bf ABSTRACT}
The renormalization of general theories with inter-family mixing of Dirac
and/or Majorana fermions is studied at the one-loop electroweak order.
The phenomenological significance of the mixing-matrix renormalization is 
discussed, within the context of models based on the SU(2)$_L\otimes$U(1)$_Y$ 
gauge group. The effect of radiative neutrino masses present in these models 
is naturally taken into account in this formulation. As an example, 
charged-lepton universality in pion decays is investigated in the 
heavy-neutrino limit. Non-decoupling heavy-neutrino effects induced by mixing 
renormalization are found to considerably affect the predictions in these 
new-physics scenarios.

\newpage
\section{Introduction}
\setcounter{equation}{0} 
\indent

Mixing effects have played a crucial r\^ole in the understanding of various 
aspects of particle phenomenology.
The most well-known examples established by experiment are the
$K^0$--$\bar{K}^0$ and $B^0$--$\bar{B}^0$ mixings \cite{pdg}, which originate
from the mixing of quarks \cite{gim}.
Furthermore, several astrophysical problems, including the
solar-neutrino-deficit puzzle, may be explained by assuming that there is also
mixing in the lepton sector leading to neutrino oscillations \cite{msw}.
In the Standard Model (SM), the mixing between the photon and the $Z$ boson is 
implemented at the loop level, and precision test are becoming sensitive to 
this effect.
In addition, many of the proposed extensions of the SM predict the possibility
of mixings between scalar, fermion, and vector fields.
In all these cases, the origin of the mixing is related to the rotation
between weak and mass eigenstates.

In order that a quantum theory yields precise quantitative predictions, it 
must be renormalizable.
The renormalization of the SM has been established \cite{taylor} and elaborated
\cite{aoki,sirlin} a long time ago.
It was noticed \cite{alberto} that the Cabibbo-Kobayashi-Maskawa (CKM)
\cite{ckm} matrix must be included in the renormalization programme as well.
However, this effect has been found to be insignificant in the SM
\cite{denner}.
The reason is that the mass differences between the down-type quarks are small 
compared to the electroweak scale, so that the CKM matrix can effectively be 
taken to be diagonal.
The situation should be very different in new-physics scenarios with large
inter-family mixings.
An attractive solution to the problem of the smallness in mass of the known
neutrinos can arise from certain SO(10) grand unified models \cite{WW} and/or
E${}_6$ superstring-inspired theories \cite{witten}, which predict
see-saw-type neutrino mass matrices with large Dirac components \cite{amon}.

The renormalization of mixing effects in extensions of the SM has not yet been 
addressed in the literature.
In this paper, we shall take the first step in this direction by elaborating 
the renormalization of general theories~\cite{SV} 
with Dirac and/or Majorana neutrinos.
Our formulation will naturally include radiative neutrino-mass contributions
\cite{BM,zpc}.
These considerations will affect a number of low-energy and LEP1/SLC
electroweak observables which have been utilized to establish bounds on the
parameter space of these models.
For illustration, we shall estimate the size of charged-lepton 
non-universality in pion decays.
Specifically, we shall consider the heavy-neutrino limit of the observable
$R_\pi= \Gamma(\pi^+\to e^+\nu )/\Gamma(\pi^+\to \mu^+\nu) $.

This paper is organized as follows.
In Section~2, we shall introduce the formalism of mass, wave-function, and
mixing-matrix renormalization in general models with Dirac and/or Majorana 
neutrinos and derive the corresponding counterterm (CT) Lagrangian.
In Section~3, we shall determine the CT's for the Dirac case in the 
on-shell renormalization scheme.
Special attention is paid to the balance of the numbers of CT's and 
renormalization conditions.
These considerations are extended to the case of Majorana neutrinos in 
Section~4.
In Section~5, we shall discuss the nature of the mixing matrices and their 
relationships in the framework of SU(2)$_L\otimes$U(1)$_Y$ theories.
The renormalization of these mixing matrices will then be performed in 
Section~6.
We shall see that those relationships among the mixing matrices 
that are enforced by the unitarity of the theory carry
over to the one-loop level, while additional identities related 
to other symmetries are violated by ultraviolet divergences.
As an application, in Section~7, we shall quantitatively analyze 
$R_\pi= \Gamma(\pi^+\to e^+\nu )/\Gamma(\pi^+\to \mu^+\nu) $ in the
heavy-neutrino limit.
Our conclusions will be summarized in Section~8.

\section{One-loop renormalization}
\setcounter{equation}{0}
\indent

We consider a general model with $N_f$ fermions, $f=(f_1,\ldots,f_{N_f})$,
which may be of Dirac and/or Majorana type.
As usual, we define the left- and right-handed components as
$f_{L,R}=\mbox{P}_{L,R}f$, where $\mbox{P}_{L,R}=(1\mp\gamma_5)/2$ are 
the corresponding chirality projection operators.
We denote the weak eigenstates and their mass matrix by a prime.
Bare parameters carry the superscript ``0". 
The bare kinetic Lagrangian in the weak basis is given by
\begin{equation}
\label{bareL}
{\cal L}^0_{kin}\ =\  i\bar{f}'^0_L\mbox{\bf1}\not\!\partial f'^0_L\, +\,
i\bar{f}'^0_R\mbox{\bf1}\not\!\partial f'^0_R\, -\,
\bar{f}'^0_L M'^0 f'^0_R\, -\, \bar{f}'^0_R M'^{0\dagger} f'^0_L ,
\end{equation}
where $\mbox{\bf1}$ is the unity matrix in flavour space and $M'^0$ is a
complex, non-diagonal mass matrix.
$M'^0$ can always be diagonalized with the help of two unitary matrices, 
$U_L^0$ and $U_R^0$, via
\begin{equation}
\label{bareM}
M^0\ =\ U^0_L M'^0 U^{0\dagger}_R\, ,
\end{equation}
where $M^0$ is a non-negative, diagonal mass matrix.
Consequently, the mass eigenstates are given by
\begin{equation}
\label{baref}
f_{L,R}^0  =\ U_{L,R}^0f'^0_{L,R}.
\end{equation}

Next, we study the renormalization of this theory. 
For this end, we write the unrenormalized parameters and fields in terms of 
renormalized ones and CT's as
\begin{eqnarray}
\label{uct}
U_{L,R}^0&=&U_{L,R}+\delta U_{L,R},\\
M'^0&=&M'+\delta M',\\
M^0&=&M+\delta M,\\
f'^0_{L,R}&=&Z'^{1/2}_{L,R}f'_{L,R}\ =\ 
\left(1+{1\over2}\delta Z'^{L,R}\right)f'_{L,R},\\
\label{flrct}
f_{L,R}^0&=&Z_{L,R}^{1/2}f_{L,R}\ =\ 
\left(1+{1\over2}\delta Z^{L,R}\right)f_{L,R}.
\end{eqnarray}
The renormalization is arranged so that the basic structure of the theory is
preserved through the order considered.
$U_{L,R}$ are still unitary, so that
\begin{equation}
\label{dUcond}
U_{L,R}\delta U_{L,R}^\dagger\ +\ \delta U_{L,R}U_{L,R}^\dagger\ =\ 0
\end{equation}
is satisfied up to terms of ${\cal O}(\delta U^2_{L,R})$.
Similarly to $M^0$, $M$ is diagonal with non-negative eigenvalues,
and Eqs.~(\ref{bareM}) and (\ref{baref}) are mapped into
\begin{eqnarray}
M&=&U_LM'U_R^\dagger\, ,\nonumber\\
f_{L,R} &=&U_{L,R} f'_{L,R}\, ,
\end{eqnarray}
respectively.
It is sufficient to consider $\delta U_{L,R}$, $\delta M$, and $Z_{L,R}^{1/2}$,
since the renormalization constants connected with the weak eigenstates are 
determined by
\begin{eqnarray}
\delta M'&=&U_L^\dagger M\delta U_R+U_L^\dagger\delta MU_R+
\delta U_L^\dagger MU_R,\\
\label{zprime}
Z'^{1/2}_{L,R}&=&U_{L,R}^{0\dagger}Z_{L,R}^{1/2}U_{L,R},
\end{eqnarray}
Equation~(\ref{zprime}) is nicely illustrated in the following diagrams:
\begin{equation}
\begin{array}{rcl}
f'^0_L & \stackrel{\displaystyle U^{0\dagger}_L}{\longleftarrow} & f^0_L\\
Z'^{1/2}_L \Big\uparrow &  & \Big\uparrow Z^{1/2}_L  \\
f'_L & \stackrel{\displaystyle U_L}{\longrightarrow } & f_L
\end{array}
\qquad\qquad\qquad
\begin{array}{rcl}
f'^0_R & \stackrel{\displaystyle U^{0\dagger}_R}{\longleftarrow} & f^0_R\\
Z'^{1/2}_R \Big\uparrow &  & \Big\uparrow Z^{1/2}_R  \\
f'_R & \stackrel{\displaystyle U_R}{\longrightarrow } & f_R
\end{array}\ .
\end{equation}

In the following, we shall work in the mass basis.
In order to find the appropriate CT Lagrangian, we eliminate the bare masses
and fields in Eq.~(\ref{bareL}),
\begin{eqnarray}
\label{L0}
{\cal L}^0_{kin} &=&  i\bar{f}_LZ^{1/2\dagger}_LZ^{1/2}_L\not\!\partial f_L\, 
+\, i\bar{f}_RZ^{1/2\dagger}_RZ^{1/2}_R\not\!\partial f_R\, -\,
\bar{f}_LZ^{1/2\dagger}_L(M+\delta M)Z^{1/2}_R f_R\nonumber\\
&& -\, \bar{f}_RZ^{1/2\dagger}_R(M+\delta M)Z^{1/2}_L f_L .
\end{eqnarray}
Up to higher-order terms, we then have
\begin{equation}
{\cal L}_{kin}^0\ =\ {\cal L}_{kin} + \delta {\cal L}_{kin},
\end{equation}
where ${\cal L}_{kin}$ is the renormalized Lagrangian and
\begin{eqnarray}
\label{deltaL}
\delta {\cal L}_{kin} &=& 
\frac{i}{2}\bar{f}_L\Big(\delta Z^L + \delta Z^{L\dagger}\Big)\not\!\partial 
f_L\, +\, 
\frac{i}{2}\bar{f}_R\Big(\delta Z^R + \delta Z^{R\dagger}\Big)
\not\!\partial f_R\\
&&-\, \bar{f}_L\Big(\delta M + \frac{1}{2} M\delta Z^R +
\frac{1}{2} \delta Z^{L\dagger} M \Big) f_R\, -\, 
\bar{f}_R\Big(\delta M + \frac{1}{2} M\delta Z^L +
\frac{1}{2} \delta Z^{R\dagger} M \Big) f_L \nonumber
\end{eqnarray}
is the CT Lagrangian.

The mixing-matrix renormalization is only important if unitary matrices 
different from unity appear in the couplings, such as the CKM matrix in the SM
charged-current interaction or the mixing matrices which give rise to
flavour-changing neutral currents in new-physics scenarios.
To elucidate this point,
we consider the charged-current interaction in a SU$(2)_L\otimes$U(1)$_Y$
model with right-handed neutrinos.  In this model, the bare Lagrangian for the
interaction of the $W$ boson with the charged leptons, $l_i$, and the
neutrinos, $n_i$, is given by
\begin{equation}
{\cal L}^0_W\ =\ -\, \frac{g^0}{\sqrt{2}}\, (W_\mu^-)^0J^{0\mu}\ +\ 
\mbox{h.c.},
\end{equation}
where $g$ is the SU(2)$_L$ coupling constant and
\begin{equation}
\label{J0mu}
J_\mu^0\ =\ \bar{l}'^0_L\mbox{\bf1}\gamma_\mu n'^0_L\ =\ 
\bar{l}^0_L B^0\gamma_\mu n^0_L\, .
\end{equation}
Here, $B^0$ is a CKM-type matrix in the lepton sector defined as
\begin{equation}
\label{B0}
B^0\ =\ U^{0,l}_L U^{0,n\dagger}_L\, ,
\end{equation} 
where $U^{0,l}_L$ and $U^{0,n}_L$ are the bare unitary
matrices that participate in the diagonalization of the 
charged-lepton and neutrino mass matrices, respectively.
Notice that the renormalization of the weak coupling $g$ and the $W$-boson 
wave-function are universal and determined by other renormalization
conditions~\cite{aoki}, so that we may separately consider
the renormalization of the charged current $J^\mu$.
Substituting Eq.~(\ref{uct}) in Eq.~(\ref{B0}), we find
\begin{equation}
\label{B0CT}
B^0\ =\ B\ +\ \delta B,
\end{equation} 
where 
\begin{equation}
\label{deltaB}
\delta B\ =\ B U^n_L\delta U^{n\dagger}_L\,+\,
\delta U^l_L U^{l\dagger}_L B\, .
\end{equation}
From Eq.~(\ref{dUcond}) we know that $U^n_L\delta U^{n\dagger}_L$ and
$\delta U^l_L U^{l\dagger}_L$ are antihermitean.
Using Eqs.~(\ref{flrct}) and (\ref{B0CT}), we can write Eq.~(\ref{J0mu}) as
$J^0_\mu = J_\mu +\delta J_\mu$, where
\begin{eqnarray}
\label{deltaJ}
\delta J_\mu &=& \bar{l}_L\, \gamma_\mu 
\Bigg\{\,\frac{1}{4} B \Big( \delta Z^{n,L} + \delta Z^{n,L\dagger}\Big)
\, +\, \frac{1}{4} \Big( \delta Z^{l,L} + \delta Z^{l,L\dagger}\Big) B
\nonumber\\
&&+B\, \bigg[\, \frac{1}{4}\Big(\delta Z^{n,L}- \delta Z^{n,L\dagger}\Big)
+ U^n_L \delta U^{n\dagger}_L \bigg]\nonumber\\
&&+\bigg[ - \frac{1}{4}\Big( \delta Z^{l,L} -\delta Z^{l,L\dagger}\Big)
+ \delta U^l_L U^{l\dagger}_L\bigg]\, B\, \Bigg\}\, n_L\, .
\end{eqnarray}
Here, we have decomposed $\delta Z^{n,L}$ and $\delta Z^{l,L}$ into hermitean
and antihermitean parts and collected all antihermitean CT's within
square brackets.
Obviously, the presence of $\delta B$ is indispensable to cancel the
ultraviolet (UV) divergences contained in the antihermitean parts of
$\delta Z^{n,L}$ and $\delta Z^{l,L}$.
This will also be illustrated in Section~7, where we shall calculate the
one-loop correction induced in $R_\pi$ by massive neutrinos.
We may fix $\delta U^n_L$ and $\delta U^l_L$ by requiring that the square
brackets in Eq.~(\ref{deltaJ}) vanish.
This leads to
\begin{eqnarray}
\label{deltaU}
\delta U^n_L &=& \frac{1}{4}\Big( \delta Z^{n,L} - \delta Z^{n,L\dagger}\Big)
U^n_L,\nonumber\\
\delta U^l_L &=& \frac{1}{4}\Big( \delta Z^{l,L} - \delta Z^{l,L\dagger}\Big)
U^l_L.
\end{eqnarray}
A similar condition was proposed in Ref.~\cite{denner} in connection with the
CKM mixing of the SM.
In order to enforce the UV finiteness of physical observables, it would be 
sufficient to replace the parentheses in Eq.~(\ref{deltaU}) with their 
UV-divergent parts, evaluated at some renormalization scale $\mu$.
This would correspond to the $\overline{\mbox{MS}}$ renormalization 
prescription.
By the same token, we might add finite, antihermitean matrices,
$c_{ij}^n$ and $c_{ij}^l$, to the terms contained within the parentheses of
Eq.~(\ref{deltaU}) and fix them by imposing some additional renormalization
conditions.
This would not affect the hermitean parts of $\delta Z^{n,L}$ and
$\delta Z^{l,L}$, and in particular their diagonal elements, which are 
arranged so that the fermion propagators have unit residues.
Although the use of Eq.~(\ref{deltaU}) is not compelling, this prescription 
seems natural and we shall adopt it in the remainder of this paper.
A detailed study of the implications of general mixing-matrix renormalization
schemes will be given elsewhere.

\section{Dirac case}
\setcounter{equation}{0}
\indent

In the following, we shall study the mass and wave-function renormalizations
in a general theory involving the mixing of $N_f$ Dirac fermions.
We denote the Dirac fermions by $f_i$, with $i=1,\ldots,N_f$.
There are two sources of imaginary contributions to the bare amplitudes.
They can arise either from the possibility of on-shell cuts through the loop
amplitudes (absorptive parts) or from complex mixing parameters
(CKM-type couplings).
Because of the hermiticity of the bare and renormalized Lagrangians,
the CT Lagrangian must be hermitean, too.
Therefore, only the dispersive parts of the one-loop amplitudes can 
participate in the renormalization procedure.
Consequently, we shall only consider the dispersive parts of the two-point 
functions in the following.
Furthermore, it will be understood that complex conjugation acts on the
coupling constants in the flavour space.

We start by considering the unrenormalized $f_j\to f_i$ transition amplitude.
Its most general form in compliance with hermiticity is\footnote{This form 
generalizes the one used in Ref.~\cite{denner}, which is specific for the SM.}
\begin{equation}
\label{unren}
\Sigma_{ij} (\not\! q)\ = \not\! q \mbox{P}_L\Sigma^L_{ij} (q^2)
 + \not\! q \mbox{P}_R\Sigma^R_{ij} (q^2) + 
\mbox{P}_L\Sigma^D_{ij}(q^2) +
\mbox{P}_R\Sigma^{D*}_{ji}(q^2), 
\end{equation}
where 
\begin{equation}
\label{herm}
\Sigma_{ij}^L(q^2) = \Sigma_{ji}^{L*}(q^2),\qquad
\Sigma_{ij}^R(q^2) = \Sigma_{ji}^{R*}(q^2).
\end{equation}
The renormalized counterparts will be denoted by a hat.
As per construction, they will satisfy relations similar to Eqs.~(\ref{unren})
and (\ref{herm}).

From the CT Lagrangian (\ref{deltaL}), we read off the relations
between the bare and renormalized transition amplitudes, {\it viz.}
\begin{eqnarray}
\label{renSL}
\hat{\Sigma}_{ij}^L (q^2) &=& \Sigma^L_{ij} (q^2)\, +\, \frac{1}{2}
\Big( \delta Z^L_{ij} + \delta Z^{L*}_{ji}\Big),\nonumber \\
\label{renSR}
\hat{\Sigma}_{ij}^R (q^2) &=& \Sigma^R_{ij} (q^2)\, +\, \frac{1}{2}
\Big( \delta Z^R_{ij} + \delta Z^{R*}_{ji}\Big),\nonumber \\
\label{renSD}
\hat{\Sigma}_{ij}^D (q^2) &=& \Sigma^D_{ij} (q^2)\, -\, \frac{1}{2}
\Big( m_i\delta Z^L_{ij} + m_j\delta Z^{R*}_{ji}\Big)\, -\, 
\delta_{ij} \delta m_i . 
\end{eqnarray}
Next, we evaluate the renormalization constants by imposing the on-shell
renormalization conditions on the renormalized transition amplitudes.
Specifically, the non-diagonal elements of $\delta Z^L_{ij}$ and
$\delta Z^R_{ij}$ are determined by requiring that 
$\hat\Sigma_{ij}(\not\! q)$ be diagonal if the external lines are put on their 
mass shells, while the diagonal elements are fixed in such a way that the
residues of the renormalized propagators are equal to unity, {\it i.e.},
\begin{eqnarray}
\label{cond1}
\hat{\Sigma}_{ij}(\not\! q) u_j(q) &=& 0,\\
\label{cond2}
\bar{u}_i(q) \hat{\Sigma}_{ij}(\not\! q) &=& 0,\\
\label{cond3}
\frac{1}{\not\! q - m_i}\hat{\Sigma}_{ii}(\not\! q) u_i(q)&=& 0,\\
\label{cond4}
\bar{u}_i(q) \hat{\Sigma}_{ii}(\not\! q) \frac{1}{\not\! q - m_i} &=& 0.
\end{eqnarray}
Conditions~(\ref{cond1})--(\ref{cond4}) imply that 
\begin{eqnarray}
\label{rc1}
m_j\hat{\Sigma}^L_{ij}(m^2_j)\, +\, \hat{\Sigma}^{D*}_{ji}(m^2_j)&=& 0,\\
\label{rc2}
m_j\hat{\Sigma}^R_{ij}(m^2_j)\, +\, \hat{\Sigma}^D_{ij}(m^2_j)&=& 0,\\
\label{rc3}
m_i\hat{\Sigma}^L_{ij}(m^2_i)\, +\, \hat{\Sigma}^D_{ij}(m^2_i)&=& 0,\\
\label{rc4}
m_i\hat{\Sigma}^R_{ij}(m^2_i)\, +\, \hat{\Sigma}^{D*}_{ji}(m^2_i)&=& 0,\\
\label{rc5}
\hat{\Sigma}^L_{ii}(m^2_i)+\hat{\Sigma}^R_{ii}(m^2_i)+
2m^2_i\Big( \hat{\Sigma}_{ii}^L{}'(m^2_i)+\hat{\Sigma}_{ii}^R{}'(m^2_i)\Big)&&
\nonumber\\
+2m_i\Big(\hat{\Sigma}_{ii}^D{}'(m^2_i)+\hat{\Sigma}_{ii}^{D*}{}'(m^2_i)\Big)
&=&0,
\end{eqnarray}
where $\Sigma'(q^2)=d\Sigma(q^2)/dq^2$. For $i\neq j$, we obtain
from Eqs.\ (\ref{rc1})--(\ref{rc4})
\begin{eqnarray}
\delta Z^L_{ij} &=& \frac{2}{m^2_i-m^2_j}\Big[ m^2_j\Sigma^L_{ij}(m^2_j)
+m_im_j\Sigma^R_{ij}(m^2_j)+m_i\Sigma^D_{ij}(m^2_j)+
m_j\Sigma^{D*}_{ji}(m^2_j)\Big],\nonumber\\
\delta Z^R_{ij} &=& \frac{2}{m^2_i-m^2_j}\Big[ m_im_j\Sigma^L_{ij}(m^2_j)
+m^2_j\Sigma^R_{ij}(m^2_j)+m_j\Sigma^D_{ij}(m^2_j)+
m_i\Sigma^{D*}_{ji}(m^2_j)\Big].
\end{eqnarray}
In the diagonal case $i=j$, the number of renormalization constants to be
determined may be reduced by exploiting the following symmetry present in the
CT Lagrangian (\ref{L0}):
\begin{equation}
\label{symm}
Z^{1/2}_{Lij} \ \to\  e^{i\theta_i} Z^{1/2}_{Lij},\qquad
Z^{1/2}_{Rij} \ \to\  e^{i\theta_i} Z^{1/2}_{Rij},
\end{equation}
where $\theta_i$ are real phases.
In this way, we may, {\it e.g.}, arrange for all $\delta Z^R_{ii}$ to be real.
Employing Eqs.\ (\ref{rc1})--(\ref{rc5}), we then obtain
\begin{eqnarray}
\label{dZLii}
\delta Z^L_{ii} &=& -\Sigma^L_{ii}(m^2_i)+
\frac{1}{m_i}\Big[\Sigma^D_{ii}(m^2_i)-\Sigma^{D*}_{ii}(m^2_i)\Big]\nonumber\\
&&-m^2_i\Big[ \Sigma^L_{ii}{}'(m^2_i)+\Sigma^R_{ii}{}'(m^2_i)\Big]
-m_i\Big[\Sigma^D_{ii}{}'(m^2_i) + \Sigma^{D*}_{ii}{}'(m^2_i)\Big],\\
\label{dZRii}
\delta Z^R_{ii} &=& -\Sigma^R_{ii}(m^2_i)
-m^2_i\Big[ \Sigma^L_{ii}{}'(m^2_i)+\Sigma^R_{ii}{}'(m^2_i)\Big]
-m_i\Big[\Sigma^D_{ii}{}'(m^2_i) + \Sigma^{D*}_{ii}{}'(m^2_i)\Big],\\
\label{dmi}
\delta m_i &=& \frac{1}{2} m_i \Big[ \Sigma^L_{ii}(m^2_i) 
+\Sigma^R_{ii}(m^2_i)\Big] +\frac{1}{2}\Big[ \Sigma^D_{ii}(m^2_i)
+\Sigma^{D*}_{ii}(m^2_i)\Big].
\end{eqnarray}

At this stage, it is instructive to investigate the mass-degenerate limit
of two fermions, $f_i$ and $f_j$. We observe that $\delta Z^L_{ij}$ and
$\delta Z^R_{ij}$ are singular in the limit $m_i\to m_j$. However,
due to condition (\ref{deltaU}), only
the hermitean parts of $\delta Z^L_{ij}$ and $\delta Z^R_{ij}$ occur
in the renormalization of physical observables. For $m_i=m_j$, 
we find
\begin{eqnarray}
\label{hermZL}
\frac{1}{2}\Big(\delta Z^L_{ij} +\delta Z^{L*}_{ji}\Big) &=&
-\Sigma^L_{ij}(m^2_j)-m^2_j\Big[ \Sigma^L_{ij}{}'(m^2_j) 
+\Sigma^R_{ij}{}'(m^2_j)\Big]\nonumber\\
&&-m_j\Big[\Sigma^D_{ij}{}'(m^2_j) + \Sigma^{D*}_{ij}{}'(m^2_j) \Big],\\
\label{hermZR}
\frac{1}{2}\Big(\delta Z^R_{ij} +\delta Z^{R*}_{ji}\Big) &=&
-\Sigma^R_{ij}(m^2_j)-m^2_j\Big[ \Sigma^L_{ij}{}'(m^2_j) 
+\Sigma^R_{ij}{}'(m^2_j)\Big]\nonumber\\
&&-m_j\Big[\Sigma^D_{ij}{}'(m^2_j) + \Sigma^{D*}_{ij}{}'(m^2_j) \Big],
\end{eqnarray}
which are indeed finite. For $i=j$, Eqs.\ (\ref{hermZL}) and (\ref{hermZR})
coincide with the hermitean parts of Eqs.\ (\ref{dZLii}) and (\ref{dZRii}),
respectively, as they should.

Finally, we verify that the number of renormalization conditions 
equals the number of CT's. To be specific, in a model with 
$N_f$ fermions, there are $4N_f$ real conditions for all $i=j$
(Eqs.~(\ref{rc1})--(\ref{rc4}) collapse to three real conditions),
and $4N_f(N_f-1)$ real conditions for all $i\ne j$
(namely Eqs.~(\ref{rc1})--(\ref{rc4})).
Thus, the total number of renormalization conditions is $4N^2_f$.
On the other hand,
counting the number of independent real CT's, we have $N_f$ mass CT's and
$4N_f^2$ wave-function renormalization constants (namely
$\delta Z^L_{ij}$, $\delta Z^R_{ij}$, and their complex conjugates).
From the number of CT's one has to subtract the $N_f$ phases,
$\theta_i$, in Eq.\ (\ref{symm}), which can be used, {\it e.g.}, to render
$\delta Z^R_{ii}$ real. As a result, the total number of independent real
CT's is $4N^2_f$ \cite{aoki}, which is equal to the one
of the real renormalization conditions.

\section{Majorana case}
\setcounter{equation}{0}
\indent

In this section, we shall study the renormalization of a general model with
$N_f$ Majorana neutrinos.
In a way, this is a generalization of the Dirac case considered in the
previous section, since a Dirac fermion may always be represented as a pair of
mass-degenerate Majorana neutrinos.
By the same token, it is possible to describe the mixing of Dirac and Majorana 
neutrinos. The kinetic Lagrangian of the model is given by 
\begin{equation}
\label{Majbare}
{\cal L}_{kin}^0 = \frac{1}{2}\Big( i\bar{f}^0_L\not\!\partial f^0_L
+i\bar{f}^0_R\not\!\partial f^0_R - \bar{f}^0_L M^0 f^0_R 
-\bar{f}^0_R M^0 f^0_L\Big).
\end{equation}
The bare and renormalized fermion fields satisfy the Majorana constraint,
\begin{equation}
\label{Majconstr}
f^0_L\ =\ (f^0_R)^C,\qquad \qquad f_L\ =\ (f_R)^C,
\end{equation}
where $C$ stands for the charge-conjugation operation.
Using Eq.~(\ref{flrct}), we thus find that
\begin{equation}
\label{MajZL*}
Z^{1/2}_L\ =\ Z^{1/2*}_R.
\end{equation}
Now, we can express the bare Lagrangian (\ref{Majbare})
in terms of renormalized quantities as
\begin{eqnarray}
\label{MLkin}
{\cal L}^0_{kin} &=& \frac{1}{2}\Big[
i\bar{f}_L Z^{1/2\dagger}_L Z^{1/2}_L \not\!\partial f_L 
+ i\bar{f}_R Z^{1/2T}_L Z^{1/2*}_L \not\!\partial f_R
-\bar{f}_L Z^{1/2\dagger}_L (M+\delta M) Z^{1/2*}_L f_R \nonumber\\
&&-\bar{f}_R Z^{1/2T}_L (M+\delta M) Z^{1/2}_L f_L \Big].
\end{eqnarray}
Decomposing this as ${\cal L}^0_{kin}={\cal L}_{kin}+\delta{\cal L}_{kin}$,
we obtain the CT Lagrangian,
\begin{eqnarray}
\label{MdLkin}
\delta {\cal L}_{kin} &=& 
\frac{i}{4}\bar{f}_L\Big( \delta Z^L + 
\delta Z^{L\dagger}\Big) \not\!\partial f_L +
\frac{i}{4}\bar{f}_R\Big( \delta Z^{L*} + 
\delta Z^{LT}\Big) \not\!\partial f_R \\
&&-\frac{1}{2}\bar{f}_L\Big( \delta M + \frac{1}{2} M \delta Z^{L*} +
\frac{1}{2}\delta Z^{L\dagger} M\Big) f_R
-\frac{1}{2}\bar{f}_R\Big( \delta M +
\frac{1}{2} M \delta Z^L + \frac{1}{2}\delta Z^{LT} M \Big) f_L. \nonumber
\end{eqnarray}

The most general form of the $f_j\to f_i$ transition amplitude between
fermionic Majorana states reads
\begin{equation}
\label{MSigma}
\Sigma_{ij} (\not\! q)\ =\ \not\! q \mbox{P}_L\Sigma^L_{ij}(q^2)
+\not\! q \mbox{P}_R\Sigma^{L*}_{ij}(q^2)+\mbox{P}_L\Sigma^M_{ij}(q^2)
+\mbox{P}_R\Sigma^{M*}_{ij}(q^2),
\end{equation}
where we have made use of the relations
\begin{equation}
\Sigma^L_{ij}(q^2)\ =\ \Sigma^{R*}_{ij}(q^2),\qquad\qquad 
\Sigma^M_{ij}(q^2)\ =\ \Sigma^M_{ji}(q^2),
\end{equation}
which follow from the Majorana condition~(\ref{Majconstr}).

From the CT Lagrangian (\ref{MdLkin}), we may read off the 
relationships between $\Sigma^L_{ij}(q^2)$, $\Sigma^M_{ij}(q^2)$,
and their renormalized counterparts, {\it viz.}
\begin{eqnarray}
\label{MrenSL}
\hat{\Sigma}^L_{ij}(q^2) &=& \Sigma^L_{ij}(q^2)\, +\, \frac{1}{2}
\Big( \delta Z^L_{ij} + \delta Z^{L*}_{ji}\Big),\\
\label{MrenSM}
\hat{\Sigma}^M_{ij}(q^2) &=& \Sigma^M_{ij}(q^2)\, -\, 
\frac{1}{2}\Big( m_i\delta Z^L_{ij} + m_j \delta Z^L_{ji}\Big)\,
-\, \delta_{ij}m_i.
\end{eqnarray}
Imposing the on-shell renormalization conditions (\ref{cond1})--(\ref{cond4}),
we obtain the renormalization constants for $i\ne j$,
\begin{equation}
\label{MdZLij}
\delta Z^L_{ij}\ =\ \frac{2}{m^2_i-m^2_j}\Big[
m^2_j\Sigma^L_{ij}(m^2_j)\, +\, m_im_j\Sigma^{L*}_{ij}(m^2_j)\, +\,
m_i\Sigma^M_{ij}(m^2_j)\, +\, m_j\Sigma_{ij}^{M*}(m^2_j)\Big],
\end{equation}
and those for $i=j$,
\begin{eqnarray}
\label{MdZLii}
\delta Z^L_{ii}& =& -\Sigma^L_{ii}(m^2_i)\, -\, 2m^2_i\Sigma^L_{ii}{}'(m^2_i)
\, -\, 
m_i\Big[ \Sigma^M_{ii}{}'(m^2_i)+\Sigma^{M*}_{ii}{}'(m^2_i)\Big]\nonumber\\
&&+\, \frac{1}{2m_i}\Big[ \Sigma^M_{ii}(m^2_i)-\Sigma^{M*}_{ii}(m^2_i)\Big],\\
\label{Mdmi}
\delta m_i &=& m_i\Sigma^L_{ii}(m^2_i)\, +\, \frac{1}{2}\Big[
\Sigma^M_{ii}(m^2_i)+\Sigma^{M*}_{ii}(m^2_i)\Big].
\end{eqnarray}
A special situation arises if a Majorana neutrino is massless at tree level.
In contrast to the SM Dirac case, this does not necessarily imply that
the mass CT $\delta m_i$ in Eq.~(\ref{Mdmi}) vanishes.
In general, it will be positive and finite, {\it i.e.}, the Majorana neutrino
receives a mass via loop effects.
Various mechanisms for generating radiative neutrino masses have been
suggested in the literature \cite{BM,zpc}; they are naturally implemented in
our formulation.

Again, we may verify that the physically relevant hermitean part of 
$\delta Z^L_{ij}$ is indeed finite in the limit $m_i\to m_j$,
\begin{eqnarray}
\frac{1}{2}\Big(\delta Z^L_{ij}+\delta Z^{L*}_{ji}\Big)& =&
-\Sigma^L_{ij}(m^2_j)-m^2_j\Big[\Sigma^L_{ij}{}'(m^2_j)+
\Sigma^{L*}_{ij}{}'(m^2_j)\Big]-m_j\Big[\Sigma^M_{ij}{}'(m^2_j)\nonumber\\
&&+\Sigma^{M*}_{ij}{}'(m^2_j)\Big].
\end{eqnarray}
Moreover, for $i=j$, we recover the hermitean part of 
Eq.\ (\ref{MdZLii}),
\begin{equation}
\frac{1}{2}\Big(\delta Z^L_{ii}+\delta Z^{L*}_{ii}\Big)\ =\ 
-\Sigma^L_{ii}(m^2_i)-2m^2_i\Sigma^L_{ii}{}'(m^2_i)
-m_i\Big[\Sigma^M_{ii}{}'(m^2_i)+\Sigma^{M*}_{ii}{}'(m^2_i)\Big].
\end{equation}
These observations reassure us of the self-consistency of our formalism.

Let us finally count the number of independent
renormalization conditions and CT's in our $N_f$-Majorana-neutrino
model.
For $i\ne j$, we have only two independent complex equations or four real
conditions resulting from Eqs. (\ref{rc1})--(\ref{rc4}),
due to the Majorana constraint (\ref{Majconstr}). 
This gives $2N_f(N_f-1)$, where we only consider
the cases $i>j$, so as to avoid double-counting 
due to the hermiticity of the renormalization conditions. In addition,
there are $3N_f$ real relations for the diagonal transitions.
So, we obtain $2N^2_f+N_f$ independent real renormalization conditions
in total. 
On the other hand, taking the Majorana constraint~(\ref{MajZL*}) into
account, we count the same number of independent CT's.
In fact, there are $2N_f^2$ wave-function renormalizations, $\delta Z^L_{ij}$
and $\delta Z^{L*}_{ij}$, and $N_f$ mass CT's.

\section{Mixing matrices in SU(2)$_L\otimes$U(1)$_Y$ theories}
\setcounter{equation}{0}
\indent

A minimal, renormalizable extension of the SM that can naturally 
accommodate heavy Majorana neutrinos is a model based on 
the SU(2)$_L\otimes$U(1)$_Y$ gauge group, in which lepton-number-violating
$\Delta L=2$ operators have been introduced in the Yukawa sector by the
inclusion of a number of $N_R$ isosinglet neutrinos. The latter 
are sometimes called right-handed neutrinos because they are blind
under SU(2)$_L$. Here, we adopt the 
conventions of Ref.~\cite{zpc} and denote the bare isosinglet weak 
eigenstates by $\nu'^0_{R_i}$ (with $i=1,\dots,N_R$).
Furthermore, we assume a number of $N_G$ weak isodoublets.
The quark sector of this model is similar to that of the minimal SM.
The bare Yukawa Lagrangian that describes the neutrino sector
reads
\begin{equation}
{\cal L}_Y^{0,\nu}\ =\ -\frac{1}{2}\Big( \bar{\nu}'^0_L,\ 
\bar{\nu}'^{0C}_R\Big)\, M'^{0,\nu}\, \left(
\begin{array}{c} \nu'^{0C}_L \\  \nu'^0_R \end{array} \right)\quad +
\quad \mbox{h.c.},
\end{equation}
where $M'^{0,\nu}$ is a complex, symmetric mass matrix of the from
\begin{eqnarray}
M'^{0,\nu}\ =\ \left( \begin{array}{cc}
0         & m'^0_D \\
m'^{0T}_D & m'^0_M \end{array} \right) .
\end{eqnarray}
It can always be diagonalized through the unitary transformation
\begin{equation}
\label{diagMnu}
U^{0,\nu T}\, M'^{0,\nu}\, U^{0,\nu}\ =\ M^{0,\nu}\, .
\end{equation}
The nonnegative, diagonal matrix $M^{0,\nu}$ contains the bare 
neutrino-mass eigenvalues. The corresponding mass eigenstates are
given by 
\begin{equation}
\label{nus}
\left( \begin{array}{c}
\nu'^0_L \\ \nu'^{0C}_R \end{array} \right)_i\ =\ 
\sum\limits^{N_G+N_R}_{j=1} U^{0,\nu *}_{ij} n^0_{Lj}, \qquad
\left( \begin{array}{c}
\nu'^{0C}_L \\ \nu'^0_R \end{array} \right)_i\ =\ 
\sum\limits^{N_G+N_R}_{j=1} U^{0,\nu}_{ij} n^0_{Rj}\, .
\end{equation}
Here, the first $N_G$ mass eigenstates, $\nu_i\equiv n_i$ ($i=1,\dots,N_G$),
are identified with the ordinary light neutrinos (if $N_G=3$),
and the remaining $N_R$ states, $N_i\equiv n_{i+N_G}$ ($i=1,\dots,N_R$), are
the new neutral leptons predicted by the model.
These neutral leptons are the heavy Majorana neutrinos,
which should be heavier than the $Z$ boson, as they have escaped 
detection in production experiments at LEP1/SLC.
The diagonalization of the charged-lepton mass matrix proceeds as outlined in
Section 2.

In our minimal model, quantum mixing effects enter via the interactions of 
the Majorana neutrinos, $n_i$, and charged leptons, $l_i$, with the 
intermediate bosons, $W^\pm$ and $Z$, and the Higgs boson, $H$.
In the `t~Hooft-Feynman gauge, the bare Lagrangians of these interactions
are~\cite{zpc}
\begin{eqnarray}
{\cal L}^0_W &=& -\ \frac{g^0}{\sqrt{2}} (W_\mu^-)^0\
\sum_{l=1}^{N_G}\sum_{j=1}^{N_G+N_R}
\bar{l}^0 \ B^0_{lj} {\gamma}^{\mu} \mbox{P}_L \ n^0_j \ + \ 
\mbox{h.c.}\ ,
\\[0.3cm]
{\cal L}^0_{G^\pm} &=& -\ \frac{g^0}{\sqrt{2}M^0_W} (G^-)^0\
\sum_{l=1}^{N_G}\sum_{j=1}^{N_G+N_R}
\bar{l}^0 \ B^0_{lj}\, \Big(m^0_l\mbox{P}_L - m^0_j\mbox{P}_R\Big) n^0_j \ + \ 
\mbox{h.c.}\ ,
\\[0.3cm]
{\cal L}^0_Z &=& -\ \frac{g^0}{4c^0_w}  Z_\mu^0\
\sum_{i,j=1}^{N_G+N_R}
\bar{n}^0_i \gamma^\mu \Big( i\Im m C^0_{ij}\ -\ \gamma_5\Re e C^0_{ij}
\Big) n^0_j\ ,\\[0.3cm]
{\cal L}^0_{G^0} &=&  \frac{i g^0}{4M^0_W}\ (G^0)^0\
\sum_{i,j=1}^{N_G+N_R}
\bar{n}^0_i \Big[ \gamma_5\, (m^0_i+m^0_j)\Re e C^0_{ij}
+\ i(m^0_j-m^0_i)\Im m C^0_{ij} \Big] n^0_j\ ,\\[0.3cm]
{\cal L}^0_H &=& -\ \frac{g^0}{4M^0_W}\ H^0\
\sum_{i,j=1}^{N_G+N_R}
\bar{n}^0_i \Big[ (m^0_i+m^0_j)\Re e C^0_{ij}
+\ i\gamma_5 (m^0_j-m^0_i)\Im m C^0_{ij} \Big] n^0_j\ ,
\end{eqnarray} 
where $G^\pm$ and $G^0$ are the charged and neutral Higgs-Kibble ghosts, 
respectively, $g$ is the SU(2)$_L$ coupling constant,
$c_w$ is the cosine of the weak mixing angle,
and $B$ and $C$ are $N_G\times (N_G+N_R)$ and $(N_G+N_R)\times(N_G+N_R)$
mixing matrices, respectively.
The bare matrices are defined as
\begin{eqnarray}
\label{defB0}
B^0_{lj}\ &=& \sum\limits_{k=1}^{N_G} V^{0,l}_{lk} U^{0,\nu\ast}_{kj}\ ,\\
\label{defC0}
C^0_{ij}\ &=&\ \sum\limits_{k=1}^{N_G}\ U^{0,\nu}_{ki}U^{0,\nu\ast}_{kj}\ .
\end{eqnarray} 
Note that $C^0_{ij}$ is hermitean.
Comparing Eqs.~(\ref{B0}) and (\ref{defB0}), we may identify
\begin{equation}
V^{0,l}\ =\ U^{0,l}_L\, , \qquad U^{0,\nu}\ =\ U^{0,n T}_L\ .
\end{equation}
Furthermore, $B^0$ and $C^0$ satisfy a number of identities, which
will turn out to be crucial to warranty the renormalizability of our model,
namely~\cite{zpc,AP,bernd}
\begin{eqnarray}
\label{ident1}
\sum\limits_{k=1}^{N_G+N_R} B^0_{l_1k}B_{l_2k}^{0\ast} =  {\delta}_{l_1l_2},\\
\label{ident2}
\sum\limits_{k=1}^{N_G+N_R} C^0_{ik}C^0_{kj} =  C^0_{ij},\\
\label{ident3}
\sum\limits_{k=1}^{N_G+N_R} B^0_{lk}C^0_{ki}  =   B^0_{li},\\
\label{ident4}
\sum\limits_{l=1}^{N_G} B_{li}^{0\ast}B^0_{lj}  =  C^0_{ij}.
\end{eqnarray}
In addition, there are identities involving the Majorana-neutrino masses,
\begin{eqnarray}
\label{identM}
\sum\limits_{k=1}^{N_G+N_R} m^0_k C^0_{ik}C^0_{jk} =  0,\qquad 
\sum\limits_{k=1}^{N_G+N_R} m^0_k B^0_{lk}C^0_{ik} =  0,\qquad 
\sum\limits_{k=1}^{N_G+N_R} m^0_k B^0_{l_1k}B^0_{l_2k} =  0.\qquad\mbox{}
\end{eqnarray} 
These relations signify the presence of lepton-number-violating interactions,
{\it e.g.}, in the possible neutrinoless double-beta decay of a nucleus
\cite{bam}.

\section{Renormalization of the mixing matrices}
\setcounter{equation}{0}
\indent

Inserting Eq.~(\ref{deltaU}) into Eq.~(\ref{deltaB}), we obtain a closed 
expression for the CT matrix of $B$,
\begin{equation}
\label{delb}
\delta B\ =\ 
\frac{1}{4}\Big(\delta Z^{l,L}\, -\, \delta Z^{l,L\dagger}\Big)\, B
-\frac{1}{4}B\, \Big(\delta Z^{n,L}\, -\, \delta Z^{n,L\dagger}\Big).
\end{equation}
By analogy, writing $C^0_{ij}=C_{ij}+\delta C_{ij}$ and requiring
that Eq.\ (\ref{ident4}) also holds true for the renormalized quantities,
{\it i.e.}, that 
\begin{equation}
\label{renB}
\sum\limits_{l=1}^{N_G} B^*_{li}B_{lj}\ =\ C_{ij}
\end{equation}
is correct to one loop, we find
\begin{equation}
\label{delc}
\delta C\ =\ 
\frac{1}{4}\Big(\delta Z^{n,L}\, -\, \delta Z^{n,L\dagger}\Big)\, C
-\frac{1}{4}C\, \Big(\delta Z^{n,L}\, -\, \delta Z^{n,L\dagger}\Big).
\end{equation}
With the help of Eqs.~(\ref{delb}) and (\ref{delc}), we may verify that the
renormalized versions of Eqs.~(\ref{ident1})--(\ref{ident3}),
\begin{equation}
\sum\limits_{k=1}^{N_G+N_R} B_{l_1k}B_{l_2k}^{\ast} =  {\delta}_{l_1l_2},\qquad
\sum\limits_{k=1}^{N_G+N_R} C_{ik}C_{kj} =  C_{ij},\qquad
\sum\limits_{k=1}^{N_G+N_R} B_{lk}C_{ki}  =   B_{li},
\end{equation}
are valid at the one-loop level.
On the other hand, the relations in Eq.~(\ref{identM}) become UV divergent
if we replace the bare parameters with their renormalized counterparts.
This may be attributed to the fact that, in contrast to
Eqs.~(\ref{ident1})--(\ref{ident4}), the relations in Eq.~(\ref{identM}) are
not enforced by unitarity.

\section{The observable $R_\pi$}
\setcounter{equation}{0}
\indent

In order to illustrate the phenomenological significance of mixing-matrix
renormalization, we shall now study possible violations of charged-current
universality in the leptonic pion decays $\pi^+\to l^+\nu$ within the
Majorana-neutrino mixing models described in Section~5.
To simplify matters, we shall consider the limit where the novel Majorana 
neutrinos are much heavier than the intermediate bosons.
We may then exploit the Goldstone-boson equivalence theorem \cite{cor} 
formulated in the `t~Hooft-Feynman gauge and include in the loop amplitudes
only the massive Higgs boson, $H$, the massless Goldstone bosons, $G^\pm$ and
$G^0$, and the heavy Majorana neutrinos, $N_i$.
In this way, we may extract the leading electroweak radiative corrections of
$O(G_Fm_{N_i}^2)$ and $O(G_FM_H^2)$, which occur because Majorana neutrinos
and the Higgs boson do not decouple in the high-mass limit.

The decay $\pi^+(P)\to l^+(k)\nu(p)$ is described by the parton-model process
$u\bar d\to l^+\nu$ in connection with the hadronic matrix element
$\langle0|\bar{d}(0)\gamma^\mu\gamma_5u(0)|\pi^+(P)\rangle=f_\pi P^\mu$,
where $f_\pi$ is the pion decay constant.
The corresponding vector-current matrix element is taken to be zero.
The tree-level transition-matrix element reads
\begin{equation}
{\cal T}_0=
-\frac{\pi\alpha}{s^2_w}f_\pi V^*_{ud}B^*_{l\nu}\frac{m_l}{M^2_W-m_\pi^2}
\bar{u}_\nu(p)\mbox{P}_Rv_l(k),
\end{equation}
where $\alpha$ is the fine-structure constant, $V$ is the CKM matrix, and
$c_w^2=1-s_w^2=M_W^2/M_Z^2$.
In a full electroweak one-loop analysis, one needs to include the corrections 
related to the $W$-boson propagator, the $Wud$ and $Wl\nu$ vertices, and the
$udl\nu$ box as well as the renormalizations of $M_W$, $\alpha$, $s_w$,
the external quark and lepton fields, and the mixing matrices, $V$ and $B$.
For the time being, we ignore photon bremsstrahlung and corrections due to
strong interactions.
Neglecting the pion, quark, and external-lepton masses, the corrected matrix
element takes the form
\begin{equation}
{\cal T}={\cal T}_0(1+\delta_{ct}^{univ}+\delta_{ct}^l)+{\cal T}_v+{\cal T}_b,
\end{equation}
where ${\cal T}_v$ and ${\cal T}_b$ are the one-particle-irreducible vertex
and box amplitudes, respectively, and we distinguish between universal and
lepton-flavour-dependent CT contributions.
To simplify the notation, we include the propagator corrections in the 
universal CT.
Specifically, we have
\begin{eqnarray}
\delta_{ct}^{univ}&=& \frac{\Pi_{WW}(0)-\delta M^2_W}{M^2_W}\, +\, 
2\frac{\delta e}{e}\, -\, 2\frac{\delta s_w}{s_w}\nonumber\\
&& +\frac{1}{V^*_{ud}} \Big[ \delta V^*_{ud}\, +\, \frac{1}{2}\Big(
\delta Z^{L*}_{d_id}V^*_{ud_i}\, +\, V^*_{u_id}\delta Z^L_{u_iu} 
\Big)\Big],\\
\label{dctl}
\delta_{ct}^l&=&\frac{1}{B^*_{l\nu}} \Big[ \delta B^*_{l\nu}\, +\, 
\frac{1}{2}\Big( \delta Z^{L*}_{n_i\nu} B^*_{ln_i}\, +\, B^*_{l_i\nu}
\delta Z^L_{l_il}\Big)\Big],
\end{eqnarray}
where summation over the quark and lepton flavours $u_i$, $d_i$, $l_i$, and 
$n_i$ is implied and \cite{aoki}
\begin{eqnarray}
\delta M^2_W &=& \Re e \Pi_{WW}(M^2_W)\, ,\nonumber\\
\frac{\delta e}{e} &=& \frac{1}{2}\Pi'_{AA}(0)\, -\, \frac{s_w}{c_w}\, 
\frac{\Pi_{ZA}(0)}{M^2_Z}\, ,\nonumber\\
\frac{\delta s_w}{s_w} & = & -\, \frac{1}{2}\, 
\frac{c^2_w}{s^2_w} \Re e \left[ \frac{\Pi_{WW}(M^2_W)}{M^2_W}
\, -\, \frac{\Pi_{ZZ}(M^2_Z)}{M^2_Z} \right].
\end{eqnarray}
The transverse gauge-boson vacuum-polarization contributions induced by the
leptons of our SU(2)$_L\otimes$U(1)$_Y$ model may be found in
Ref.~\cite{bernd}.
The calculation considerably simplifies in the heavy-neutrino limit,
$m_{N_i}\gg M_W$.
According to the equivalence theorem, it is then sufficient to compute the
vertex and lepton-mixing diagrams depicted in Fig.~1, while ${\cal T}_b=0$.
Furthermore, we may project out the relevant vertex form factor by
\begin{equation}
\label{dvl}
{\cal T}_0^*{\cal T}_v=|{\cal T}_0|^2(1+\delta_v^l).
\end{equation}

The present knowledge of the radiative corrections to
$\Gamma(\pi^+\to l^+\nu)$ in the SM has been nicely summarized by Marciano and
Sirlin \cite{MS}.
The short-distance corrections and most uncertainties cancel in the ratio
$R_\pi=\Gamma(\pi^+\to e^+\nu)/\Gamma(\pi^+\to \mu^+\nu)$.
Apart from helicity-suppressed terms of
$O[(\alpha/\pi)(m_\mu^2/m_\rho^2)\ln(m_\rho^2/m_\mu^2)]$, where the
typical hadronic mass scale $m_\rho$ is used as a demarcation between short- 
and long-distance loop corrections, only QED corrections survive.
The latter consist of a pointlike-pion contribution \cite{SMB} and a
structure-dependent bremsstrahlung correction, which is suppressed by
$m_\pi^4/m_\rho^4$.
The leading term of the pointlike-pion contribution is given by
$-(3\alpha/2\pi)\ln(m_\mu^2/m_e^2)\approx-3.7\%$ and may be summed via the
renormalization group; it makes up the bulk of the SM correction to $R_\pi$.
According to Ref.~\cite{MS}, the current SM prediction is
\begin{equation}
\label{rsm}
R^{SM}_\pi\ =\ (1.2352\pm 0.0005)\times 10^{-4},
\end{equation}
where the error is mainly due to model dependence.
As we shall see in the following, the presence of heavy Majorana neutrinos
might be manifested by a significant shift in the theoretical prediction of
$R_\pi$ relative to $R_\pi^{SM}$.
In turn, confrontation of the modified prediction with experiment, together
with similar analyses for other low-energy and LEP1/SLC observables, will
allow one to improve the constraints on the parameter space of the
Majorana-neutrino models under consideration.
In our approximation, the one-loop-corrected value of $R_\pi$ in the
Majorana-neutrino scenario is given by
\begin{equation}
\label{rpi}
R_\pi=R_\pi^0\, (1+C),
\end{equation}
where $R_\pi^0$ emerges from the SM tree-level result by multiplication with
$|B_{e\nu}|^2/|B_{\mu\nu}|^2$ and
\begin{equation}
\label{rc}
C=2\Re e(\delta_v^e+\delta_{ct}^e-\delta_v^\mu-\delta_{ct}^\mu).
\end{equation}
The universal contribution $\delta_{ct}^{univ}$ has dropped out.
Analytic results for $\delta_v^l$ and $\delta_{ct}^l$ are listed in the 
Appendix.
We note that $\delta_{ct}^{univ}$ and $\delta_v^l+\delta_{ct}^l$ are 
separately UV finite.
The significance of mixing-matrix renormalization becomes apparent by 
observing that $\delta_v^l+\delta_{ct}^l$ contains the UV-divergent
contribution from $\delta B_{l\nu}^*$.
We may refine Eq.~(\ref{rpi}) by including the SM radiative corrections of
Ref.~\cite{MS}.
For this end, we write $R_\pi^0=(|B_{e\nu}|^2/|B_{\mu\nu}|^2)R_\pi^{SM}$,
where $R_\pi^{SM}$ is given by Eq.~(\ref{rsm}).

For clarity, we shall henceforth consider a very simple realization of a
SU(2)$_L\otimes$U(1)$_Y$ model with Majorana neutrinos, which is of
phenomenological interest and displays the essential features of mixing
renormalization.
We shall assume that the SM is extended by two heavy Majorana neutrinos, $N_1$
and $N_2$, which only mix with the leptons of one family, the first one, say.
The other two families are assumed to be standard.
In the notation of Section~5, this corresponds to $N_G=1$ and $N_R=2$.
The relevant mixing-matrix elements are determined by solving
Eqs.~(\ref{ident1})--(\ref{identM}), with the result that
\begin{eqnarray}
|B_{e\nu}|^2 &=& 1-(s^{\nu_e}_L)^2, \qquad
B_{eN_1}\ =\ \frac{\rho^{1/4}s^{\nu_e}_L}{\sqrt{1+\rho^{1/2}}},\qquad
B_{eN_2}\ =\ i\rho^{-1/4}B_{eN_1},\\
C_{N_1N_1} &=& \frac{\rho^{1/2}(s^{\nu_e}_L)^2}{1+\rho^{1/2}},\qquad
C_{N_2N_2} \ =\ \rho^{-1/2} C_{N_1N_1},\qquad
C_{N_1N_2}\ =\ -C_{N_2N_1}\ =\ i\rho^{-1/4} C_{N_1N_1}\, ,\nonumber
\end{eqnarray}
where $\rho=m^2_{N_2}/m^2_{N_1}$ and $s^{\nu_e}_L$ measures the degree of
light-heavy neutrino mixing \cite{LL}.
Without loss of generality, we may assume that $\rho\ge1$.
As mentioned above, we set $s^{\nu_\mu}_L=s^{\nu_\tau}_L=0$.
Then, Eq.~(\ref{rc}) becomes $C=2\Re e(\delta_v^e+\delta_{ct}^e)$.
This pattern of mixing may be motivated by the non-observation of the
decay $\mu\to e\gamma$ or the absence of $\mu$--$e$ conversion in nuclei.
Furthermore, a global analysis of low-energy data gives 
the following upper limits \cite{BGKLM}:
\begin{equation}
\label{bounds}
(s^{\nu_e}_L)^2\ <\ 0.010\, ,\qquad (s^{\nu_\mu}_L)^2\ <\ 0.0020
\end{equation}   
at the 95\% confidence level.
We should note that electroweak corrections, mainly those originating from the
SM, are taken into account only for a limited number of low-energy observables, 
such as the muon decay width, in the evaluation of the bounds given in
Eq.~(\ref{bounds}).
Therefore, the so-derived bounds should be considered to be of tree-level
accuracy.

The results for $\delta_v^l$ and $\delta_{ct}^l$ given in the Appendix are 
valid for $M_H$ arbitrary as long as $m_{N_1},m_{N_2}\gg M_W$.
If we are interested in the non-decoupling effects of $O(G_Fm_{N_i}^2)$, we
may put $M_H=0$.
In this limit, we have
\begin{equation}
\label{corr}
C=-\frac{G_F(s^{\nu_e}_L)^4m_{N_2}^2}{4\pi^2\sqrt2(1+\sqrt\rho)^2}
\left[3+\left(1-\frac{1}{\sqrt\rho}\right)f(\rho)
+(1-\sqrt\rho)f\left(\frac{1}{\rho}\right)\right],
\end{equation}
where
\begin{equation}
f(x)=\frac{1}{2}\left(1-\frac{1}{x^2}\right)\ln|1-x|-\frac{1}{4}-\frac{1}{2x}.
\end{equation}
Notice that, to the order considered, we may introduce $G_F$ in
Eq.~(\ref{rpi}) by substituting $G_F=(\pi\alpha/\sqrt2s_w^2M_W^2)$;
the adjustment proportional to $\Delta r$ \cite{sirlin} will only appear at
the two-loop order because the tree-level expression for $R_\pi$ does not
contain $\alpha$.
The correction $C$ is throughout negative, takes on the simple form
\begin{equation}
\label{unitrho}
C=-\frac{3G_F(s^{\nu_e}_L)^4m_{N_1}^2}{16\pi^2\sqrt2}
\end{equation}
for $\rho=1$, and exhibits the limiting behaviour
\begin{equation}
\label{bigrho}
C=-\frac{G_F(s^{\nu_e}_L)^4m_{N_1}^2}{8\pi^2\sqrt2}
\left[\left(1-\frac{1}{\sqrt\rho}\right)\ln\rho+\frac{11}{2}
+\frac{7}{6\sqrt\rho}+O\left(\frac{1}{\rho}\right)\right]
\end{equation}
for $\rho\gg1$.
It is interesting to observe that $C$ is of $O(G_Fm_{N_1}^2)$, {\it i.e.}, it 
scales with the squared mass of the lighter neutrino.

We are now in a position to estimate how the value of $s_L^{\nu_e}$ extracted
from the measurement of $R_\pi$ is affected by the inclusion of loop effects 
due to heavy Majorana neutrinos.
If we neglect these loop effects, we obtain the value $(s_{L0}^{\nu_e})^2$ by
equating $R_\pi^{exp}=R_\pi^{SM}[1-(s_{L0}^{\nu_e})^2]$.
On the other hand, including these effects, we extract the corrected value
$(s_L^{\nu_e})^2=(s_{L0}^{\nu_e})^2+\Delta(s_L^{\nu_e})^2$ from
$R_\pi^{exp}=R_\pi^{SM}[1-(s_L^{\nu_e})^2](1+C)$, where $C$ is evaluated with
$s_L^{\nu_e}$.
Consequently, the relative shift in $(s_L^{\nu_e})^2$ is
\begin{equation}
\frac{\Delta(s_L^{\nu_e})^2}{(s_L^{\nu_e})^2}=
\frac{1-(s_L^{\nu_e})^2}{(s_L^{\nu_e})^2}C
\approx\frac{1}{(s_L^{\nu_e})^2}\left(\frac{R_\pi}{R_\pi^0}-1\right).
\end{equation}
In the degenerate case, $m_{N_1}=m_{N_2}=m_N$, we then have
\begin{equation}
\label{estimate}
\frac{\Delta(s_L^{\nu_e})^2}{(s_L^{\nu_e})^2}\approx
-16\%\times\left(\frac{s_L^{\nu_e}m_N}{1~\mbox{TeV}}\right)^2.
\end{equation}
Inserting in Eq.~(\ref{estimate}) the perturbative upper limit \cite{CFH}
\begin{equation}
s^{\nu_e}_L m_N\approx m_D\simlt1~\mbox{TeV},
\end{equation}
we find that $(s_L^{\nu_e})^2$ is reduced by 16\%.

With the help of Eq.\ (\ref{exact}), we shall now investigate the dependence of
$-\Delta(s_L^{\nu_e})^2/(s_L^{\nu_e})^2\approx
(1-R_\pi/R_\pi^{SM})/(s_L^{\nu_e})^2$ on $m_{N_1}$, $m_{N_2}$, and $M_H$.
We first consider the degenerate case, with $\rho=1$.
In Fig.~2, we show $-\Delta(s_L^{\nu_e})^2/(s_L^{\nu_e})^2$ as a function of
$m_N$ for $M_H=100$, 300, 500, and 1000~GeV, assuming (a)
$(s_L^{\nu_e})^2=0.01$ and (b) 0.007.
We observe that, at low values of $m_N$, the relative reduction of
$(s_L^{\nu_e})^2$ strongly depends on $M_H$; it increases by almost one order
of magnitude as $M_H$ runs from 100~GeV to 1~TeV.
Comparing Figs.~2(a) and (b), we see that, for $m_N$ in the TeV range, the
relative reduction of $(s_L^{\nu_e})^2$ is slightly smaller for the lower value
of $(s_L^{\nu_e})^2$.
The high-$m_N$ regime is well described by Eq.~(\ref{unitrho}).
Next, we study the non-degenerate case for $M_H=200$~GeV.
In Fig.~3, we display the $\rho$ dependence of
$-\Delta(s^{\nu_e}_L)^2/(s^{\nu_e}_L)^2$ for $m_{N_1}=0.5$, 1, 2, and 5~GeV,
again assuming (a) $(s_L^{\nu_e})^2=0.01$ and (b) 0.007.
We see that, in compliance with Eq.~(\ref{bigrho}),
$-\Delta(s^{\nu_e}_L)^2/(s^{\nu_e}_L)^2$ grows quadratically with $m_{N_1}$
for $\rho$ fixed, while it grows logarithmically with $\rho$ for $m_{N_1}$
fixed.
From Fig.~3(a), we read off a 20\% effect for $m_{N_1}=5$~TeV and
$m_{N_2}=50$~TeV.
On the experimental side, the measurements at the Tri-University Meson 
Facility (TRIUMF) \cite{TRIUMF} and the Paul Scherrer Institute (PSI)
\cite{PSI} yielded
\begin{eqnarray}
R^{exp}_\pi (\mbox{TRIUMF}) 
&=& (1.2265\pm 0.0034(\mbox{stat})\pm 0.0044(\mbox{syst})) \times 10^{-4},
\nonumber\\
R^{exp}_\pi (\mbox{PSI})
&=& (1.2346\pm 0.0035(\mbox{stat})\pm 0.0036(\mbox{syst})) \times 10^{-4},
\end{eqnarray}
respectively.
This represents a remarkable reduction in error, by a factor of 3, relative to
the previous value \cite{bryman}, $R^{exp}_\pi=(1.218\pm 0.014)\times 10^{-4}$. 
It is therefore reasonable to expect that, in the near future, experiments
will become sensitive to 10\% effects in $(s_L^{\nu_e})^2$.

\section{Conclusions}
\indent

The renormalization of general theories with inter-family 
mixing between fermionic Dirac and/or Majorana states was 
studied to one loop in the electroweak on-shell scheme.
Special attention was paid to the renormalization of the mixing matrix
\cite{alberto}, which plays a central r\^ole in such theories.
Similarly to the renormalization prescription for the CKM matrix of the SM
proposed in Ref.~\cite{denner}, we adjusted the mixing-matrix CT's in such
a way that they precisely cancel the antihermitean parts of the wave-function
renormalization constants.
Our formulation naturally takes possible radiative-neutrino-mass contributions
into account.

The phenomenological implications of mixing renormalization 
for Majorana-neutrino mass models with large SU(2)$_L$-breaking Dirac
components~\cite{AP} were analyzed in Section~7.
In such scenarios, low-energy observables may receive sizeable corrections due
to the non-decoupling of heavy neutrinos.
As an example, the electroweak corrections to the observable $R_\pi$
were estimated in the heavy-neutrino limit.
It was found that they may reduce the tree-level values of $R_\pi$ predicted
in these models by up to 0.2\%.
This has to be contrasted with the present theoretical error in the SM
prediction, which is $\pm0.04\%$ \cite{MS}.
Future experiments to be performed at TRIUMF and PSI may be sensitive to such
new-physics phenomena.

Finally, we wish to emphasize that our formalism for the mixing renormalization
of fermionic states may straightforwardly be extended to theories
which involve the mixing of scalar or vector particles, such as the 
mixing of scalar quarks in supersymmetric theories.

\vspace{2cm}
\noindent
{\bf Acknowledgements.}
We thank Ansgar Denner for useful comments regarding Ref.~\cite{denner}.
AP is grateful to the Werner-Heisenberg-Institut for the kind hospitality
extended to him during a visit when part of this work was performed.

\newpage

\def\theequation{\Alph{section}.\arabic{equation}}
\begin{appendix}
\setcounter{equation}{0}
\section{Analytic expressions}
\indent

In this appendix, we list the lepton transition amplitudes as well as the
corrections $\delta_v^l$ and $\delta_{ct}^l$ defined in Eqs.~(\ref{dctl}) and
(\ref{dvl}), respectively, to one loop in the Majorana-neutrino models of
Section~5 assuming $m_{N_i}\gg M_W$ and $M_H$ arbitrary.
We adopt the Passarino-Veltman \cite{pv} conventions for the standard one-loop
integrals in dimensional regularization, implemented with the Minkowskian
metric, $g^{\mu\nu}=\mbox{diag}(1,-1,-1,-1)$, as in Appendix~A of
Ref.~\cite{pr}.

The charged-lepton and Majorana-neutrino mixing amplitudes depicted in
Figs.~1(b) and (c) are found to be
\begin{eqnarray}
\Sigma^l_{ij}(\not\! p) &=& -\frac{\alpha}{8\pi s_w^2M^2_W}\, 
B_{l_ik}B^*_{l_jk}\, \Big[ \not\! p \Big( m^2_k\mbox{P}_L +
m_{l_i}m_{l_j}\mbox{P}_R\Big)\, B_1 (p^2,m^2_k,M^2_W)\nonumber\\
&&+\Big( m_{l_i}\mbox{P}_L + m_{l_j}\mbox{P}_R\Big) m^2_k
B_0(p^2,m^2_k,M^2_W)\Big],\nonumber\\
\Sigma^n_{ij}(\not\! p) &=& -\frac{\alpha}{16\pi s_w^2M^2_W}\, 
\bigg\{ \Big[ \not\! p \mbox{P}_L\Big( m_i C^*_{ik} +
m_k C_{ik}\Big)\Big( m_k C_{kj} + m_j C^*_{kj} \Big)\nonumber\\
&&+\not\! p \mbox{P}_R \Big( m_i C_{ik} + m_k C^*_{ik}\Big)
\Big( m_k C^*_{kj} + m_j C_{kj} \Big)\Big]\Big[ B_1(p^2,m^2_k,M^2_Z)
+B_1(p^2,m^2_k,M^2_H) \Big] \nonumber\\
&&+\Big[ \mbox{P}_L m_k \Big( m_i C_{ik} + m_k C^*_{ik} \Big)
\Big( m_k C_{kj} + m_j C^*_{kj} \Big)+\mbox{P}_R m_k 
\Big( m_i C^*_{ik} + m_k C_{ik} \Big)\nonumber\\
&&\times\Big( m_k C^*_{kj} + m_j C_{kj} \Big)\Big]
\Big[B_0(p^2,m^2_k,M^2_Z)-B_0(p^2,m^2_k,M^2_H)\Big] \bigg\},
\end{eqnarray}
respectively.
Here and in the following, the Majorana indices $n_i$ are abbreviated by $i$, 
and it is summed over the indices of the heavy Majorana neutrinos.
For the $Wl\nu$ vertex correction due to Fig.~1(a) and the
lepton-flavour-dependent CT contribution to $\Gamma(\pi^+\to l^+\nu)$ we find
\begin{eqnarray}
\label{exact}
\delta_v^l&=&\frac{\alpha}{8\pi s_w^2M_W^2}C_{ii}m_i^2
\Bigl[C_{24}(0,0,0,M_H^2,m_i^2,0)+C_{24}(0,0,0,0,m_i^2,0)\Bigr],
\nonumber\\
\delta_{ct}^l&=&\frac{\alpha}{32\pi s_w^2M_W^2}\biggl\{
C_{ii}m_i^2\Bigl[B_1(0,m_i^2,M_H^2)+3B_1(0,m_i^2,0)\Bigr]\nonumber\\
&&+|C_{ij}|^2m_j^2\Bigl[B_1(m_i^2,m_j^2,M_H^2)+B_1(m_i^2,m_j^2,0)
-B_1(0,m_j^2,M_H^2)-B_1(0,m_j^2,0)\nonumber\\
&&-B_0(m_i^2,m_j^2,M_H^2)+B_0(m_i^2,m_j^2,0)+B_0(0,m_j^2,M_H^2)-B_0(0,m_j^2,0)
\Bigr]\nonumber\\
&&+C_{ij}^2m_im_j\Bigl[B_1(m_i^2,m_j^2,M_H^2)+B_1(m_i^2,m_j^2,0)
-\frac{m_j^2}{m_i^2}\Bigl(B_0(m_i^2,m_j^2,M_H^2)\nonumber\\
&&-B_0(m_i^2,m_j^2,0)-B_0(0,m_j^2,M_H^2)+B_0(0,m_j^2,0)\Bigr)\Bigr]\biggr\}.
\end{eqnarray}

\end{appendix}

\newpage

\centerline{\Large\bf Figure Captions}

\newcounter{fig}
\begin{list}{\bf\rm Fig.~\arabic{fig}: }{\usecounter{fig}
\labelwidth1.6cm \leftmargin2.5cm \labelsep0.4cm \itemsep0ex plus0.2ex }

\item Feynman graphs contributing to the observable 
$R_\pi=\Gamma (\pi^+ \to e^+\nu)/\Gamma (\pi^+ \to \mu^+\nu)$ in 
the heavy-neutrino limit:
{\bf (a)} $Wl\nu$ vertex corrections,
{\bf (b)} charged-lepton transition amplitudes,
{\bf (c)} neutrino transition amplitudes.

\item $-\Delta (s^{\nu_e}_L)^2/(s^{\nu_e}_L)^2\approx
(1-R_\pi / R^0_\pi ) / (s^{\nu_e}_L)^2$ as a function of
$m_N=m_{N_1}=m_{N_2}$ for $M_H=100$, 300, 500, 1000~GeV, assuming
$(s^{\nu_\mu}_L)^2=0$ and {\bf (a)} $(s^{\nu_e}_L)^2=0.01$ or {\bf (b)}
$(s^{\nu_e}_L)^2=0.007$.

\item $-\Delta (s^{\nu_e}_L)^2/(s^{\nu_e}_L)^2\approx
(1-R_\pi / R^0_\pi ) / (s^{\nu_e}_L)^2$ as a function of
$\rho=m^2_{N_2}/m^2_{N_1}$ for $m_{N_1}=0.5$, 1, 2, 5~TeV, 
assuming $M_H=200$~GeV, $(s^{\nu_\mu}_L)^2=0$,
and {\bf (a)} $(s^{\nu_e}_L)^2=0.01$ or {\bf (b)} $(s^{\nu_e}_L)^2=0.007$.

\end{list}

\end{document}